\documentstyle[aps,prd,preprint,tighten]{revtex}

\newcommand{\quantity}[1]{\overstar{#1}}
\newcommand{\lengtheffect}{\Upsilon}

\begin{document}
\draft
\title{Propagation-Based General Relativity}
\author{Edward M. Schaefer\thanks{E-mail: schaefer@plansys.com}}
\address{Planning Systems Inc., 7923 Jones Branch Dr., McLean, VA 22102}
\date{September 6, 1996}
\maketitle
\begin{abstract}
It is assumed that the radial propagation of light with respect to the naive
coordinate system of the observer is uniform and isotropic and that the
physical rate of propagation of light is the same for all observers.  In
accelerated frames of reference, these assumptions lead to the findings that
the measured value of $c$ is a function of the gravitational energy per unit
mass (GEPUM) of the observer, and that this is due to the physical
characteristics of the standard measuring-devices being a function of their
GEPUM.  The consequences of these findings include observers who at rest
with respect to each other assigning different values to the same physical
separation, the mixed metric tensor ${g^\mu}_\nu$ describing how gravitation
affects measuring-devices, and the De-Broglie wavelength being a function of
an object's GEPUM.  How the measured values of various types of physical
quantities are affected is described.  The gravitational Doppler shifting of
light is viewed differently.  The correct value for the deflection of light
is obtained without the use of Huyghen's Principle.  The Schwarzschild
solution is re-examined: The physical size of radial coordinate $2m$ is 0, a
traveler must perceive himself to go an infinite distance to reach the
radial coordinate of $2m$, and gravitational self-potential energy reaches a
minimal value at the radial coordinate $3m$.  Therefore, black holes do not
exist in this theory; Gravitational collapse forms hyper-massive star-like
objects instead.
\end{abstract}
\pacs{04.20.Cv, 04.40.Dg, 04.70.-s, 98.62.Ai, 98.63.Js}
\section*{Introduction}

The cornerstone of the theories of Relativity is the statement that the
propagation of light is uniform and isotropic for all observers.  The term
``uniform'' usually refers to Einstein's assumption that all observers
measure the same value for the speed of light
($c$)\cite{Einstein-SR,Einstein-1911}.  This assertion is based on two
additional assumptions:
\begin{list}{$\bullet$}{\setlength{\topsep}{0ex plus0.2ex}
			\setlength{\parsep}{0ex plus0.2ex}
			\setlength{\itemsep}{0ex plus0.2ex}}
\item
$c$ is physically the same for all observers, which will be referred to
as Absolute Uniformity.  Absolute Uniformity is the heart and soul of
Relativity, and is retained in the theory proposed in this article.  
\item
All measuring devices of identical construction are physically identical.
This will be referred to as Construction Uniformity.
\end{list}

In Special Relativity (SR), another form of uniformity involving the
propagation of light exists: The radial\footnote{This qualifier is not
needed in flat space-times.  However, the distorting effects of curved
space-times can make the rate of lateral propagation of light appear to be
different than that of its radial propagation.  The resulting loss of
isotropy would invalidate the resulting theory if these lateral effects were
treated as being physically as perceived.} propagation of light with
respect to the naive coordinate system of every observer is instantaneously
the same throughout the universe, where the naive coordinate system is that
created by the observer when it is assumed that light travels in straight
lines.  This will be referred to as Extended Uniformity.

Although Extended Uniformity and Construction Uniformity peacefully co-exist
in SR, the same does not apply to General Relativity (GR).  This is because
the coordinate size of a standard measuring-rod and the rate at which a
standard clock ticks both decrease as their gravitational energy per unit
mass (GEPUM) decreases.  When Absolute Uniformity and Construction
Uniformity are assumed, it is predicted that the rate of propagation of
light with respect to a naive coordinate system will be diminished as it
descends into a gravitational field\cite{Einstein-1911}, thereby violating
Extended Uniformity.  The resulting theory, which is currently accepted, is
referred to in this article as Measurement-Based GR (MGR).

On the other hand, when Extended Uniformity is used instead of Construction
Uniformity, the GR effects on the coordinate features of clocks and
measuring-rods produce the prediction that $c$ increases as one's GEPUM
decreases.  If Absolute Uniformity is also assumed, then it cannot be a
change in the physical quantity which is the rate of propagation of light
($\quantity{c}$) which causes the change in $c$.  Instead, the physical
attributes of the standards of length and time must now be a function of
their GEPUM and not solely a function of their construction.

In this article, it is assumed that Absolute Uniformity and Extended
Uniformity are the underlying principles on which the universe operates.
The theory constructed on these postulates is referred to as
Propagation-based GR (PGR).  In this theory, physical space, time, mass, and
charge are as measured with respect to the naive coordinate system of an
observer with a given GEPUM.  The resulting model of the universe makes a
number of surprising predictions, including a finding that black holes can
not exist, and that gravitational collapse halts itself before a singularity
can be formed.

\section*{Fundamentals}
\subsection{Quantities, Values, and Effects}
\label{sec:buzzwords}

Given the loss of Construction Uniformity, how should standards of
measurement be created and/or reconciled?  In PGR, the physical standards
continue to be described by their construction because:
\begin{list}{$\bullet$}{\setlength{\topsep}{0ex plus0.2ex}
			\setlength{\parsep}{0ex plus0.2ex}
			\setlength{\itemsep}{0ex plus0.2ex}}
\item
We are physical beings and what we experience is a physical world.  The
physical standards therefore should help a physical being to describe the
various phenomena of the universe as perceived by itself.
\item
The existing tensors of GR already assume the use of measuring-devices which
are of identical construction, and permit the measurements made with them in
different frames of reference to be reconciled.  To go with another type of
standard would require the development of a new mathematical framework to
support it.
\end{list}

The physical size of the standard measuring-devices being a function of
their GEPUM means that different observers may assign different values to
the same physical quantity.  Not only does this principle apply to $c$, but
to all velocities as well as the measurement of spatial and temporal
distances, mass\footnote{The effect of a change in GEPUM on the
standard mass has been documented by Einstein\cite{Einstein-1911}.},
acceleration\footnote{This includes the measurement of gravity by different
observers in a uniformly accelerated box.}, etc.  As a result one must
differentiate between a physical \emph{quantity} and the \emph{value} an
arbitrary observer measures it to have.  In this article, this is done
symbolically by placing a star over a normal symbol to represent a quantity
(such as $\quantity{c}$), while using the normal symbol to represent its
measured value (such as ${c}$).  For example: For two observers $K$ and $K'$
in an accelerated system both MGR and PGR agree that $\quantity{c}' =
\quantity{c}$.  However, in MGR $c' = c$, while that is not true in PGR.

The quantity notation will be used when it does not matter which observer is
measuring the quantities in the expression.  In that case, what matters is
having those quantities measured in a consistent fashion.  For example, the
PGR expression for determining how much faster a clock at a different
potential than ones own will run in a short accelerated box is\footnote{This
may be derived from Eq. (1) (in \S\ref{sec:metric-tensor}).}  $\sqrt {1 +
2\quantity{g} \quantity{z} /\quantity{c}^2}$, where $\quantity{g}$ is the
quantity of acceleration throughout the box, and $\quantity{z}$ is the
quantity of distance in the direction opposite that of gravity in the box
between ones own position and that of the clock being examined.  Who
measures $g$, $z$, and $c$ is irrelevant.  Since the result is
dimensionless, all of the length and time variances cancel out.  Instead,
what matters is that all of these quantities be measured by a given observer
while that observer has a given GEPUM.

In many cases, the difference between a quantity and its value is trivial.
For a single observer whose position in a stable accelerated system is not
changing, or who is making all of their measurements at a given instant in
time, the measured value of a quantity is a reasonable description of the
quantity itself.  Another example is $E = mc^2$: There is no practical
difference between it and $\quantity{E} = \quantity{m}{\quantity{c}}^2$.  On
the other hand, it is quantities that are subject to the conservation laws
of physics.  (The consequences of this observation will manifest themselves
in \S\ref{sec:collapse}).

In PGR, gravitation affects measured values in two ways: When two observers
obtain different values for the same quantity, that is called an
\emph{observer-based effect}, since what has changed is the observer's units
of measure as opposed to the quantity being measured.

On the other hand, consider the case of an observer with two identical
atomic clocks who sends one of them to another gravitational potential.
After the transfer, the observer will find that the clock which was moved is
no longer keeping the same time as that of the clock which he kept.  This is
because in the process of being transfered, the quantity of time which is a
standard unit of time came to be different for the clock which was moved.
This is called an \emph{object-based effect}, since it is caused by the
object being measured having a different GEPUM.

The physical constants of nature may be either quantities or values.  Those
that describe the ``construction'' of the universe such as $c$, Plank's
Constant ($h$), etc. are \emph{constant quantities}; while those that which
describe the construction of ordinary matter such as the quantum of
electrical charge ($e$) and the mass of the electron ($m_e$) are locally
measured values or \emph{local constants}.  The difference is important
since constant quantities are not subject to object-based effects, while the
quantities described by local constants are.

\subsection{Effect on the metric tensor \boldmath $g_{\mu\nu}$}
\label{sec:metric-tensor}

Extended Uniformity is not inconsistent with either the Equivalence
Principle or general covariance.  As a result, The basic tensor mathematics
of GR are retained in PGR, including the Einstein Field Equations.  However,
the interpretation of the components of the metric tensor of spacetime
$g_{\mu\nu}$ is modified in PGR.

First of all, the components of the mixed metric tensor ${g^\mu}_\nu$
describe how the physical properties of the standard clock and measuring-rod
are modified by a change of GEPUM.  For example, ${g^\mu}_\nu$ expressed in
terms of the Newtonian potential $\quantity{\Phi}$ for $\quantity{\Phi} < 0$
is\cite{Ohanian-pot}:
\begin{mathletters}
\label{pot-metric}
\begin{eqnarray}
\label{pot-time}
{g^0}_0 &=& 1 + 2\quantity{\Phi}/\quantity{c}^2, \\
\label{pot-space}
{g^1}_1 = {g^2}_2 = {g^3}_3 &=& (1 + 2\quantity{\Phi}/\quantity{c}^2) 
^{-1}, \\
{g^\mu}_\nu &=& 0\text{,  }\mu \neq \nu .
\end{eqnarray}
Eq.\ (\ref{pot-time}) indicates that a clock ticks $(1 + 2\quantity{\Phi}
/\quantity{c}^2)^{1/2}$ as fast as one at the base potential, and 
Eq.\ (\ref{pot-space}) indicates that spatial distances as measured in the
$x$, $y$, and $z$ directions from a position at potential $\quantity{\Phi}$
will be $(1 + 2\quantity{\Phi} /\quantity{c}^2)^{-1/2}$ as large as those
measured from the base potential.
\end{mathletters}

Secondly, $g_{\mu\nu}$ continues to describe how to reconcile the views
from the various frames of reference.  This use is more important given
that the coordinate values now have direct physical significance.
Therefore, the transformation rule for physical measurements of space and
time between observers $K$ and $K'$ in an accelerated system is given
in PGR by:
\begin{equation}
\label{PGR-trans}
x'^{\mu} \,=\, {a^\mu}_\alpha \sqrt{{g^\alpha}_\beta {(g^{-1})^\beta}_\nu}
\,x^{\nu},
\end{equation}
where ${a^\mu}_\alpha$ is the Lorenz tensor, ${g'^\alpha}_\beta$ is the
mixed metric tensor for observer $K'$, and ${(g^{-1})^\beta}_{\nu}$ is
the inverse mixed metric tensor for observer $K$.  This replaces the MGR
rule of using measuring-rods at rest in the lattice to measure distance
with.

Thirdly, the expression for the invariant spacetime interval $ds$ is still
given by $ds^2 \,=\,g_{\mu\nu} dx^{\mu}dx^{\nu}$.  The $dx^{\mu}$ and
$dx^{\nu}$ continue to be locally measured values.

Finally, ones view of spacetime may be distorted by gravitational lensing
and the overall curvature of spacetime.  In this case, there are occasions
where the field equations predict that the lateral effects of gravitation
on the measurement of space and time are not the same as the radial
effects.  The radial values then describe the direct physical effects of
gravitation, while the variance between the radial and lateral effects
describe the form and magnitude of the distortions caused by the
curvature of spacetime.

As was previously mentioned, $c$ is not the same for all observers in PGR.
Under Extended Uniformity, $c$ is given by:
\begin{equation}
\label{c-value}
c \:=\: c_0 \sqrt{\sum_{\nu=1}^3 \frac{{g^{1}}_{\nu}}{{g^0}_0}},
\end{equation}
where $c_0$ is the value of $c$ at some base potential where ${g^\mu}_\nu =
\delta^\mu_\nu$ by definition\footnote{This is not to be confused with the
SR case, where ${g^\mu}_\nu = \delta^\mu_\nu$ for all of spacetime.},
$\delta^\mu_\nu$ is the Kronecker delta, and $x^1$ is the direction of the
radial propagation of light from a position at the base potential.  Eq.\
(\ref{c-value}) indicates that all observers will measure the same value for
$c$ if $\sum_{\nu=1}^3 {g^1}_\nu = {{g^0}_0}$, but this is not normally the
case.  In fact, it is usual for $\sum_{\nu=1}^3 {g^1}_\nu = 1/{{g^0}_0}$,
which results in Eq.\ (\ref{c-value}) taking on the form:
\begin{equation}
\label{lightspeed-2}
c = c_0 / {g^0}_0 .
\end{equation}

\section*{General Results}
\subsection{The observer-based effects of gravitation on measured values}
\label{sec:quantities}

In this section, the effects of an observer's GEPUM on the measured value of
various types of quantities is studied. This is done by associating the
values for quantities as measured by observers $K$ and $K'$ who are at rest
with respect to each other in an accelerated system, and using standards for
time, length, mass-energy, and electrical charge which are of identical
construction.  Each effect will be expressed in the form:
\[q' ~=~ q\lengtheffect^f , \] 
where $q$ is a quantity as measured by observer $K$, $q'$ is that same
quantity as measured by observer $K'$, $\lengtheffect$ is the factor for the
expansion in the measurement of length with Eqs.\ (\ref{pot-metric}) and
(\ref{PGR-trans}) indicating that $\lengtheffect = [{g'^0}_0{(g^{-1})^0}_0]
^{-1/2}$, and $f$ is an arbitrary factor.  Several of these effects have
been previously documented, but in a different context\cite{Bowler}.

A complete list of the observer-based effects is given in Table
\ref{table:effects}.  How the entries in this table were obtained is
described below.

By definition, the measurement of lengths using measuring-rods of identical
construction is modified by the relationship:
\begin{equation}\label{length-eff-def}
l' = l\lengtheffect .
\end{equation}
For clocks of identical construction, it is shown by Eq.\ (\ref{pot-metric})
that the time effect is given by:
\[ 
t' = t\lengtheffect^{-1} .
\] 
These fundamental observer-based effects (and the others that will be
described in this section) may be combined to create other observer-based
effects using the rules for powers.  For example, velocity is given by $v =
l/t$;  Therefore $v' = l'/t' = (l/t)\lengtheffect^{[1 - (-1)]} =
v\lengtheffect^2$.

Mass-energy raises a problem.  A certain amount of matter as could be used
as our standard for either mass or energy.  It is known that as a given
quantity of matter is loses GEPUM, its mass-energy as measured by an
observer with a given GEPUM is decreased by its loss of gravitational
energy\cite{Einstein-1911}.  Given a standard mass with a mass-energy of
$\cal M$ when it is at rest at the base potential, this decrease occurs in
such a way that its mass-energy when it is at rest at another potential
($\cal M'$) is given by\cite{Wald-mass}:
\begin{equation}\label{mass-mod}
{\cal M' = M} \sqrt{g_{00}} \equiv {\cal M}\lengtheffect^{-1}.
\end{equation}
If the rest energy in the standard mass $E$ is taken to be a fundamental
unit, then Eq.\ (\ref{mass-mod}) produces a fundamental
observer-based effect of:
\begin{equation}\label{energy-fund}
E' = E\lengtheffect.
\end{equation}
On the other hand, rest mass of the standard mass $m$ could be taken to
be a fundamental unit.  In this case, Eq.\ (\ref{mass-mod}) produces a
fundamental observer-based effect of:
\begin{equation}\label{mass-fund}
m' = m\lengtheffect .
\end{equation}

Eqs.\ (\ref{energy-fund}) and\ (\ref{mass-fund}) cannot both be assumed.
This is due to the mass-energy relationship $E=mc^2$.  Since $c$ varies by
$c' = c\lengtheffect^2$, if the rest energy in a standard mass with the
observer's GEPUM is treated as a fundamental unit, then as the observer's
GEPUM changes the measured rest mass of the standard mass $M$ is modified by
the relationship:
\begin{equation}\label{local-mass}
M' = M\lengtheffect^{-4}.
\end{equation}
On the other hand, if the rest mass of standard mass is used as a
fundamental unit, then the measured rest energy associated with it
when it has the observer's GEPUM $\cal E$ is modified as the observer's
GEPUM changes by the relationship:
\begin{equation}\label{local-energy}
{\cal E' = E} \lengtheffect^4.
\end{equation}
Because of this conflict Table \ref{table:effects} contains two columns of
observer-based effects: one for the mass-fundamental case and another for
the energy-fundamental case.

Eqs.\ (\ref{local-mass}) and\ (\ref{local-energy}) are combinations of an
observer-based effect and an object-based one.  For example: if the
observer-based effect for energy in Eq.\ (\ref{energy-fund}) is taken to be
fundamental, then for the measured value $m'$ of an absolute quantity of
mass the observer-based relationship
\begin{equation}\label{mass-quantity-effect}
m' = E'/c'^2 = (E/c)\lengtheffect^{(1 - 2*2)} = m\lengtheffect^{-3}
\end{equation}
will be obtained. The object-based effect on mass-energy given in Eq.\
(\ref{mass-mod}) is also in effect since both the observer and the object
have been transitioned to the other potential.  It is the product of the
effects in Eqs.\ (\ref{mass-quantity-effect}) and\ (\ref{mass-mod}) that
produces the effect in Eq.\ (\ref{local-mass}).

What happens to an electrical charge as its GEPUM changes?  If all electrons
are of identical construction and the electrical force is mediated by
photons, then the absolute electrical flux per unit time for the electron
decreases as its GEPUM decreases due to time dilation.  Therefore, the
standard electrical charge $C$ is subject to an object-based effect of:
\begin{mathletters}
\begin{equation}\label{charge-obj-effect}
C' = C\lengtheffect^{-1},
\end{equation}
and electrical charge itself ${\cal C}$ is subject to a fundamental
observer-based effect of:
\begin{equation}\label{charge-obs-effect}
{\cal C' = C} \lengtheffect.
\end{equation}
\end{mathletters}

Constant quantities such as the electrical permeability of the vacuum
$\epsilon_0$ and the magnetic permittivity of the vacuum $\mu_0$ can not be
subject to object-based effects, but like $c$, $\epsilon_0$ and $\mu_0$ may
be subject to an observer-based effect.  Suppose that observer $K$ is
measuring $\epsilon_0$ by noting the force $\quantity{F}$ that two charges
of strength $\quantity{\cal C}_1$ and $\quantity{\cal C}_2$ with a stable
GEPUM and constant separation of $\quantity{r}$ exert on each other.  As
observer $K$'s GEPUM changes, Eq.\ (\ref{length-eff-def}) calls for the
measured distance between the charges $r$ to change while Eq.\
(\ref{charge-obs-effect}) calls for the measured values of the charges
(${\cal C}_1$ and ${\cal C}_2$) to change in a similar manner and the
energy-fundamental effects listed in Table\ \ref{table:effects} indicate
that they are measured to exert the same force $F$ on each other.  Putting
these considerations together and using the relationship $\epsilon_0 = {\cal
C}_1 {\cal C}_2 / Fr^2$, an energy-fundamental observer-based
effect of:
\[ 
\epsilon_0' = \epsilon_0
\] 
is obtained. For magnetism, since $c$ is subject to an observer-based effect
of $c' = c\lengtheffect^2$ and $c^2 = 1/\epsilon_0\mu_0$, $\mu_0$ is subject
to an energy-fundamental observer-based effect of:
\[\mu_0' = \mu_0\lengtheffect^{-4} .\]

On the other hand, the use of mass fundamental standards produces the
opposite result: $\mu_0' = \mu_0$ and $\epsilon_0' = \epsilon_0
\lengtheffect ^{-4}$.  This is due to the mass-fundamental observer-based
effect on force being $F' = F\lengtheffect^4$ while the measurement of force
is uniform in the energy-fundamental case ($F' = F$).

\subsection{The gravitational Doppler shifting of light}
\label{sec:red-shift}

The gravitational Doppler shifting of light provides a way of illustrating
the observer-based effects and the differences between MGR and PGR.  Under
both theories, the rate at which an atom with a lower GEPUM is observed to
vibrate is less than that of an atom with the observer's GEPUM, and that
explains the change in frequency.  However, the two theories make different
predictions as to how the wavelength of the light behaves.

\newcommand{\gravpoteffsq}{(1 + 2\quantity{\Phi}/\quantity{c}^2)}

Suppose that a photon emitted in a given atomic transition by an atom with
the GEPUM of a base observer (observer A).  Observer A will measure a
frequency of $\nu$ for the photon and a corresponding wavelength of $\lambda
= c/\nu$.  The atom is now moved to another location which is at a
gravitational potential of $\quantity{\Phi}$ with respect to that of
observer A and kept at rest there.  From Eq.\ (\ref{pot-metric}),
$\lengtheffect = \gravpoteffsq^{-1/2}$ for an observer at the new potential
(observer B).

The same atomic transition now produces a photon which is measured by
observer B to have a frequency $\nu' = \nu$. In MGR observer B will also
observe the photon's wavelength $\lambda'$ as being $\lambda' = \lambda$
because Construction Uniformity demands that observer B's value for the
speed of light $c'$ be $c$.  At the time the photon is emitted, its
wavelength in observer A's frame of reference is now given by two
considerations: Because of time dilation, the photon's frequency is $\nu'' =
\nu \lengtheffect^{-1}$, while the coordinate speed of light with respect to
observer A at the position of observer B is $c'' = c \lengtheffect^{-2}$;
resulting in a coordinate wavelength of $\lambda'' = c'' / \nu'' = \lambda
\lengtheffect ^ {-1}$.  When the photon finally reaches observer A, the
coordinate speed of light will be $c''' = c$ while the frequency of the
photon will be $\nu''' = \nu''$, resulting in the wavelength finally
becoming
\begin{equation}\label{red-shift}
\lambda''' = c''' / \nu''' = c / \nu'' = \lambda \lengtheffect .
\end{equation}

Under PGR, the situation changes: To the observer B, the photon's frequency
is still\footnote{This may not be exactly true in PGR due to its predicted
effects on the fine structure constant as described in 
\S\ref{sec:quantum-effects}.  However, it is usable as a first approximation
since the frequency shift from the gravitational Doppler shift will be much
greater than that from the gravitational fine-structure effects.}  $\nu'' =
\nu$, but because $c' = c \lengtheffect^2$ a wavelength of $\lambda ' =
\lambda \lengtheffect^2$ is observed.  Due to Extended Uniformity, the
light propagates at constant speed with respect to all observers, and
because of that its wavelength can not change.  At first glance, this is a
problem, since it is expected to have the wavelength given in Eq.\
(\ref{red-shift}) when it is reaches observer A.  However, observer A
measures lengths to be $\lengtheffect^{-1}$ as long as those measured by
observer B.  Therefore, in the frame of reference of observer A, $\lambda ''
= \lambda ''' = \lambda \lengtheffect$.

This exercise indicates that the light was already Doppler shifted when it
was emitted.

\subsection{The deflection of light}
\label{sec:deflection}

The deflection of light in a gravitational field has been verified, but
Einstein's explanation for this deflection, Huyghen's
Principle\cite{Einstein-1911,Einstein-GR}, can not be used under Extended
Uniformity.  In PGR, the deflection of light is solely a consequence of the
constantly changing state of motion which defines being accelerated.  The
PGR prediction for the magnitude of the bending of light is given below.
This prediction produces the same result as that obtained using tensor
mathematics\cite{Ohanian-deflection}, thereby validating that tensor
mathematics may be used in PGR.  However, this method is not usable in MGR
because its treatment of physical distance is improper when Construction
Uniformity is being assumed\cite{Rindler}.

To compute the magnitude of the bending, suppose that a photon enters
an accelerated box at time $\quantity{t}=0$, that the box is of width
$\quantity{w}$, and that the acceleration of gravity in the $-z$ direction
in the box is $\quantity{g}$.  The angle of the deflection of light $\theta$
is given by the relationship $\theta = \quantity{c}_z / \quantity{c}$, where
$\quantity{c}_z$ is the downward velocity of light obtained as it propagates
across the box, and $\quantity{c}_z \ll \quantity{c}$.  Therefore, in terms
of time, $\theta$ can be calculated using:
\begin{equation}\label{deflection}
\theta = \frac{1}{\quantity{c}} \int_0^{\quantity{t}_\gamma} 
\quantity{g}\,d\quantity{t} ,
\end{equation}
where $\quantity{t}_\gamma = \quantity{w} / \quantity{c}$.

Since $\quantity{c}$ is a constant speed, Eq.\ (\ref{deflection}) can be
represented as a space-based integral.  However, in making this conversion,
one must consider that while $\int \quantity{g}\, d\quantity{t}
\to \quantity{v}$, at the same time $\int \quantity{g}\, d\quantity{x} \to
\quantity{v}^2 / 2$.  Eq.\ (\ref{deflection}) is also generalized 
for all directions of propagation by replacing $\quantity{g}$ with $\partial
\quantity{\Phi} / \partial \quantity{z}$.  The resultant equation is:
\begin{equation}
\theta = \frac{2}{\quantity{c}^2} \int_0^\quantity{w} \frac{\partial \quantity{\Phi}}{\partial \quantity{z}} d\quantity{x},
\end{equation}
where $\quantity{x}$ is the initial direction of propagation of the light
and $\quantity{z}$ is a direction perpendicular to $\quantity{x}$ for which
the deflection is being measured.

\subsection{Quantum Mechanics and the Bohr Radius}
\label{sec:quantum-effects}

Combining Quantum Mechanics (QM) and PGR results in changes being made to
QM.  The basic reason for this involves the De Broglie wavelength and the
way it is affected by an object's GEPUM.

Take an object at the position of observer $K$ which he measures to have a
rest mass of $m$ and a velocity of $v$.  This object has a momentum of $p$ and
a De Broglie wavelength $\lambda$ of:
\begin{equation}\label{De-Broglie}
\lambda = h/p,
\end{equation}
where $h$ is Plank's Constant.  The object is then transfered to the
position of observer $K'$ and is made to move with that same velocity of $v$ as
measured by observer $K'$.  The object's rest mass as measured by observer
$K$ will now be given by the object-oriented relationship $m' = m
\lengtheffect^{-1}$, and its velocity as measured by observer $K$ will then be
$v' = v \lengtheffect^{-2}$.  Therefore, the object's momentum as measured
by observer $K$ is:
\begin{equation}\label{momentum-trans}
p' = p\lengtheffect^{-3}. 
\end{equation}

Since $\lambda$ is a length, it is expected to be governed by the
object-oriented relationship
\begin{equation}\label{wavelength-trans}
\lambda ' = \lambda \lengtheffect^{-1}.
\end{equation}
However, since $h$ is a constant quantity, Eqs.\ (\ref{De-Broglie}) and\
(\ref{momentum-trans}) predict that $\lambda ' = \lambda \lengtheffect^3$,
which does not agree with Eq.\ (\ref{wavelength-trans}). To get the
De~Broglie wavelength to come out right, it is postulated that it is
affected by an object's GEPUM in such a way that
\begin{equation}\label{De-Broglie-PGR}
\lambda = h\lengtheffect^{-4}/p .
\end{equation}

To demonstrate that Eq.\ (\ref{De-Broglie-PGR}) holds in PGR, the Bohr radius
of the atom $a_0$ is examined.  The equation for the Bohr radius is:
\begin{equation}\label{Bohr-radius}
a_0 = 4 \pi \epsilon_0 h^2 / m_e e^2,
\end{equation}
where $m_e$ is the mass of the electron and $e$ is the charge of the
electron.  The constant quantities $\epsilon_0$ and $h$ do not change for
observer $K$ when the atom is moved to the position of observer $K'$, while
Eq.\ (\ref{mass-mod}) indicates that the local constant $m_e$ is subject to
an object-based effect of $\lengtheffect^{-1}$, and Eq.\
(\ref{charge-obj-effect}) indicates that the local constant $e$ is subject
to an identical object-based effect.  Because of this, Eq.\
(\ref{Bohr-radius}) calls for an object-based effect of $a_0 ' = a_0
\lengtheffect^3$ instead of the desired $a_0 ' = a_0 \lengtheffect^{-1}$.
It is no coincidence that this is the same problem that we had with the De
Broglie wavelength.  Eq.\ (\ref{Bohr-radius}) is the radius at which the
wavelength of the electron is exactly adequate to wrap around a proton once,
and therefore it implicitly assumes that Eq.\ (\ref{De-Broglie}) holds.
Because of this, PGR calls for Eq.\ (\ref{Bohr-radius}) to be rewritten as:
\begin{equation}\label{Bohr-radius2}
a_0 = 4 \pi \epsilon_0 h^2 \lengtheffect^{-4} / m_e e^2 .
\end{equation}

The changes which produced Eq.\ (\ref{Bohr-radius2}) affect the binding
energy of the atom $\cal E$, which is given by ${\cal E} = -e^2 / \epsilon_0
a_0$ for a Hydrogen atom in its ground state.  The object-based effects
$e' = e \lengtheffect^{-1}$ and $a_0 ' = a_0 \lengtheffect^{-1}$ predict
that ${\cal E}' = {\cal E} \lengtheffect^{-1}$.  This is consistent with the
finding in \S\ref{sec:red-shift} that the photons emitted by atoms with lower
GEPUMs were already of a lower energy when they were emitted:  The atom
itself has a lower binding energy when it has a lower GEPUM.

Another effect from the merger of PGR and QM involves the fine structure
\emph{factor}\footnote{This is usually called the fine-structure
\emph{constant}, but as this paragraph shows, it is not expected to be the
same for all atoms in PGR.} $\alpha$, which is given by $\alpha = e^2 /
\epsilon_0 hc$.  When observing the fine-structure factor of an atom with
another GEPUM $\alpha'$, PGR's object-based effects indicate that the
quantity which is $e$ for that atom will be different [$e' = e\lengtheffect
^ {-1}$ from Eq.\ (\ref{charge-obj-effect})] while $\epsilon_0,\; h\text{,
and } c$ are constant quantities, resulting in an object-based effect on the
fine structure factor of:
\begin{equation}\label{fine-struc-eff}
\alpha' = \alpha \lengtheffect^{-2} .
\end{equation}
For an observer with the same GEPUM as this atom, PGR's observer-based
effects (see Table \ref{table:effects}) indicate that the local value of the
product $\epsilon_0 h$ is uniform while $c$ is modified ($c' =
c\lengtheffect ^2$) and $e$ is a local constant.  This once again produces
$\alpha' = \alpha \lengtheffect^{-2}$.  For an observer with an arbitrary
GEPUM measuring the fine-structure factor $\alpha''$ of an atom at the base
potential, the observer-based effect on charge [$e' = e \lengtheffect$ from
Eq.\ (\ref{charge-obs-effect})] kicks in, and combining the square of this
effect with the effect in Eq.\ (\ref{fine-struc-eff}) produces a result of
$\alpha'' = \alpha$.  It may be possible to astronomically confirm the
existence of the object-based effect on $\alpha$ as noted in Appendix
\ref{app:exper}.

\section* {Massive objects}
\subsection{The Schwarzschild solution}
\label{sec:schwarzschild}

The Schwarzschild solution is an exact external solution to the Einstein
field equations for a spherically symmetric, non-rotating, massive object
(which is referred to as a Schwarzschild object).  To derive it, one starts
with an equation for space-time intervals around a Schwarzschild
object\cite{Ohanian-Schwarzchild} which, in PGR terms, is:
\begin{equation}\label{Schwarzschild-orig}
ds^2 = A(\quantity{r})d\quantity{t}^2 - B(\quantity{r})(\quantity{r}^2\,
d\theta^2 + \quantity{r}^2 \sin^2 \theta\, d \phi^2 ) -
C(\quantity{r})d\quantity{r}^2 ,
\end{equation}
where $s$ is a spatial-temporal interval, $\quantity{r}$ is the quantity of
distance from the center of the massive object (CMO), $\quantity{t}$ is the
temporal quantity, $\theta$ and $\phi$ are spherical surface coordinates,
and $A(\quantity{r})$, $B(\quantity{r})$, and $C(\quantity{r})$ are
arbitrary functions.

To simplify the math for the Schwarzschild equation, a transformation
is done on the radial coordinate by defining a new radial coordinate $r'$
which is: $r' \equiv \quantity{r} \sqrt {B(\quantity{r})}$ , and which
turns Eq.\ (\ref{Schwarzschild-orig}) into:
\begin{equation}\label{Schwarzschild-local}
ds^2 = A'(r')dt^2 - B'(r')dr'^2 
- r'^2 d \theta^2 - r'^2 \sin^2 \theta\, d \phi^2 ,
\end{equation}
where $t$ is the value of $\quantity{t}$ as measured from $r'$.

What is the physical interpretation of $r'$ in PGR?  To find the answer,
first consider that $\sqrt{B(\quantity{r})}$ is the factor by which a
measuring-rod oriented perpendicular the the line running between itself and
the CMO will be shortened with respect to that of some base observer.  Since
such shortenings are based on GEPUM, the same shortening holds for a ruler
oriented parallel to the line running between itself and the CMO.  Then take
the case of a base observer located at an arbitrary $\quantity{r}$ from the
CMO, who measures their own distance from the CMO to be $r_0$, and who is at
rest with respect to the massive object.  Since $B(\quantity{r}) \equiv 1$
for the base observer by definition, $r' = r_0$ at the position of the base
observer.  Since the location of the base observer is arbitrary, it may be
conjectured that $r'$ is the distance that an observer at any $\quantity{r}$
from the CMO and at rest with respect to it will measure for itself.  This
conjecture is confirmed by considering the case of the $r'$ for a position
that the base observer measures to be at distance $r_1$ from the CMO.  Since
$\sqrt{B(r_1)}$ is the factor by which distances as measured from $r_1$
differ from distances as measured at $r_0$, $r_1 '$ is the distance that an
observer at $\quantity{r}_1$ from the CMO and at rest with respect to the
massive object will measure as being their own distance from the CMO.

Further steps (which are not discussed here) result in $A'(r')$ and
$B'(r')$ being:
\begin{mathletters}
\label{Schwarzschild-factors}
\begin{eqnarray}
A'(r') & = & 1 - C/r' , \\
B'(r') & = & 1/(1 - C/r') ,
\end{eqnarray}
where $C$ is a constant of integration.  To be the same for all observers,
$C$ must be a quantity as measured from some base potential.  This requires
that the base potential of the Schwarzschild equation be determined.  As
stated above, at the base potential $g_{\mu\nu} = \eta_{\mu\nu}$.  In Eqs.\
(\ref{Schwarzschild-local}) and (\ref{Schwarzschild-factors}), this can only
be true for extremely large values of $r'$.  Therefore, the base potential
is infinitely distant from the massive object.
\end{mathletters}

When the value of $C$ is determined, it is found that $C = 2m$,
where $m = M G/ c^2$, $M$ is the mass of the gravitating
object as measured by a distant observer, $G$ is the gravitational
constant as measured by a distant observer, and $c$ is as measured
by a distant observer.

In MGR, it is usual to remove the prime from $r'$.  This is not done in PGR
because $r'$ and $r$ are physically distinct in this theory.  $r$ is radial
distance as measured by some base observer (in this case a very distant
observer), while $r'$ is the self-measured radial distance for the local
observer.  How then are $r$ and $r'$ related?  Since the base observer's
measuring-rod is $\sqrt{B'(r')}$ a big as that of the local observer,
\begin{equation}\label{radius-loc2dist}
r = r'/\sqrt{B'(r')} = \sqrt{r' (r' - 2m)} .
\end{equation}
The inverse equation for Eq.\ (\ref{radius-loc2dist}) is:
\begin{equation}\label{radius-dist2loc}
    r' = m + \sqrt {r^2 + m^2} .
\end{equation}

Eqs.\ (\ref{radius-loc2dist}) and (\ref{radius-dist2loc}) show that a radial
coordinate of $r' = 2m$ corresponds to $r = 0$.  This indicates that a black
hole can not be embedded in the naive coordinate system of an outside
observer.  This holds true in MGR\cite{lawson} as well as PGR, but since
naive coordinate systems measures physical reality in PGR, the inability to
embed it in the coordinate system means that the black hole can not exist.
In addition:
\begin{list}{$\bullet$}{\setlength{\topsep}{0ex plus0.2ex}
			\setlength{\parsep}{0ex plus0.2ex}
			\setlength{\itemsep}{0ex plus0.2ex}}
\item
$\lengtheffect \equiv B'(r')$.  Therefore at $r' = 2m$, $\lengtheffect =
\infty$.  This indicates that the Schwarzschild radius is measured locally
with a measuring-rod which is \emph{physically of zero length}.
\item
At the surface of a point object, a Newtonian potential of $\Phi =
-\quantity{c}^2 / 2$ exists.  This is consistent with SR since it means that
no free-falling object can be made to go faster than $\quantity{c}$ with
respect to any other frame of reference.
\item
In Newtonian physics, an infinite amount of energy is needed to escape from
the ``surface'' of a point mass.  Similarly, in GR an infinite amount of
energy is needed to escape from $r' = 2m$\cite{Ohanian-energy}.  In PGR,
these two views of gravitational collapse become one and the same.
\end{list}

How many measuring-rods of identical construction at rest with respect to
a massive object and on a line going through its center are needed to
bridge the gap between two radial coordinates $r_1'$ and $r_2'$?
This is given by:
\begin{equation}\label{rods-basic}
\int_{r_1'}^{r_2'} \sqrt{{g^1}_1}\, dx^1 ,
\end{equation}
where $dx^1$ is an increment of distance in the radial direction and
${g^1}_1 \equiv B'(r')$.  In MGR, Eq.\ (\ref{rods-basic}) is converted into:
\begin{equation}\label{rods-MGR}
\int_{r_1'}^{r_2'}\sqrt{1/(1 - 2m / r')}\, dr' .
\end{equation}
However, in the PGR viewpoint, there is a problem with 
Eq.\ (\ref{rods-MGR}): The shortening of a unit measuring-rod to being
$[B'(r')]^{-1/2}$ units long occurs with respect to the units of a distant
observer, and that observer measures distances using $r$ instead of $r'$.
Therefore, in PGR Eq.\ (\ref{rods-basic}) is converted into:
\begin{equation}\label{rods-PGR}
\int_{r_1'}^{r_2'}\sqrt{1/(1 - 2m / r')}\, dr .
\end{equation}

To solve Eq.\ (\ref{rods-PGR}), $1/(1 - 2m/r')$ is first modified to
$r'/(r' - 2m)$.  Then using Eq.\ (\ref{radius-dist2loc}) to convert $r'$
into $r$ produces:
\begin{equation}\label{rods-int1}
\sqrt {r'/(r' - 2m)} = \sqrt {\frac{m + \sqrt {r^2 + m^2}}
{-m + \sqrt {r^2 + m^2}}} = \frac{m + \sqrt {r^2 + m^2}}{r} .
\end{equation}
Substituting Eq.\ (\ref{rods-int1}) into the integral from Eq.\
(\ref{rods-PGR}) produces:
\begin{equation}\label{rods-int2} 
\int \frac{m + \sqrt {r^2 + m^2}}{r} dr = \sqrt {r^2 + m^2} 
+ m \ln \left( \frac {r^2}{m + \sqrt{r^2 + m^2}} \right) + C ,
\end{equation}
where C is a constant of integration.  Eq.\ (\ref{radius-loc2dist}) is then
used to represent Eq.\ (\ref{rods-int2}) in term of $r'$:
\[ 
\sqrt {r^2 + m^2} + m \ln \biglb( r^2 / (m + \sqrt {m^2 + r^2}) \bigrb) 
+ C = r' - m + m \ln (r' - 2m) + C .
\]
Therefore:
\begin{equation}\label{Schwarzschild-travel}
\int_{r_1'}^{r_2'} \sqrt {1/(1 - 2m / r')}\, dr  =
   r_2' - r_1' + m \ln \biglb( (r_2' - 2m) / (r_1' - 2m) \bigrb) .
\end{equation}

At radial coordinates $r' \gg 2m$, Eq.\ (\ref{Schwarzschild-travel}) reduces
to $r_2' - r_1'$, which is normal for unaccelerated frames of reference.
But as one approaches $r' = 2m$, the term $m \ln \biglb( (r_2' - 2m) / (r_1'
- 2m) \bigrb)$ comes to dominate.  This term has a peculiar effect: To
travel from a radial coordinate of $r_1' = 2m +a$ to $r_2' = 2m + a/2$, one
must travel for a locally perceived distance of at least $m \ln (1/2) =
-.693 m$.  Furthermore this applies to every decrease of the difference
between one's radial coordinate and $2m$ by $\case{1}{2}$.  This is called
the Achilles Effect, because it is reminiscent of Zeno's Paradox.

\subsection{Gravitational Collapse}
\label {sec:collapse}

Because $r' = 2m$ describes a point object in PGR, black holes can not exist,
and the Achilles Effect [Eq.\ (\ref{Schwarzschild-travel})] makes this
singularity unreachable.  However, these considerations do not eliminate
the possibility of infinite gravitational collapse.  It will now be shown
that PGR does not permit that.  The key concept in disallowing infinite
gravitational collapse is that it is \emph{quantities} which are conserved,
and that locally measured values will not describe physical quantities in a
consistent manner.  Instead, quantities must be measured from a base frame
of reference.

Take the case of a massive object which is a hollow sphere of infinitesimal
thickness. All of its mass is at the same radial coordinate $r'$ and
therefore it all has the same GEPUM.  Eq.\ (\ref{mass-mod}) indicates that
after an object with an initial rest mass of $M$ falls though a potential of
$\Phi$, its final rest mass will be $M' = M\lengtheffect^{-1}$ where, in the
vicinity of a Schwarzschild object,
\[ 
\lengtheffect \equiv B'(r') = \sqrt {r'/(r' - 2m)} .
\] 

The potential energy of self-gravitation is a function of the radius of
the object and the square of its mass.  This results in the quantity for
the potential energy of self-gravitation ${\Psi}$ of the hollow
sphere as measured by a distant observer being given by:
\begin{equation}\label{sphere-energy}
\Psi = -\frac{M^2 G}{r} \lengtheffect^{-2} 
= -\frac{M^2 G}{\sqrt {r'(r' - 2m)}} \frac{r' - 2{m}}{r'}
= -{M}^2 {G} \sqrt {\frac{r' - 2m}{r'^3}} .
\end{equation}

There are two situations where Eq.\ (\ref{sphere-energy}) goes to 0: At
$r' = \infty$ due to $\infty$ being the denominator, and $r' = 2m$ due to
the numerator going to 0.  (The case of $r' = \infty$ represents the mass
being spread out across the whole universe, while the case of $r' = 2m$
represents the matter in the object as having lost all of its rest mass.)
Therefore, there exists a radius at which gravitational self-potential
energy has a minimal value.  At that radius $d \Psi /dr' = 0$.

To solve this equation, $\Psi^2 = -M^4G^2(r' - 2m) / r'^3$ will be used so
that is is now $d \Psi^2 / dr' = 0$ which is being solved.  Ignoring the
$-M^4G^2$ term (since it is constant), this produces:
\begin{equation}\label{min-pot-rad}
   \frac{d \Psi^2}{dr'}
= \frac{d}{dr'} \left ( \frac{1}{r'^2} - \frac{2m}{r'^3} \right ) 
= -\frac{2}{r'^3} + \frac{6m}{r'^4} = 0,~ r' = 3m .
\end{equation}
So at $r' = 3m$, an object's potential energy of self-gravitation is at its
minimal value.  A radial coordinate of $r' = 3m$ corresponds to a Newtonian
potential of $\quantity{\Phi} = -\quantity{c}^2 / 3$.  This means that an
object which is too massive to be stabilized by quantum electrostatic forces
will be stabilized by gravity in such a way that its mass exists as an
average Newtonian potential of $\quantity{\Phi} = -\quantity{c} ^2 / 3$.

It is fair to note that a collapsing object will retain its kinetic energy
of collapse, and may therefore collapse to a Newtonian potential of
$\quantity{\Phi} < -\quantity{c}^2 / 3$.  In this case, the Achilles
effect comes into play to halt the collapse.  As one approaches the limiting
potential of $\quantity{\Phi} = -\quantity{c}^2 / 2$, the locally measured
radius of the object will become almost constant at something slightly
greater than $2m$, but the effective depth of the object [which is obtained
from Eq.\ (\ref{Schwarzschild-travel})] will begin to increase dramatically.
The result is a large increase in the effective volume at a given potential.
The collapsing object will therefore begin to spread out within that
effective volume.  Eventually, the increase in volume will allow quantum
electrostatic forces to reassert themselves, and the atoms being put into
excited states will absorb the kinetic energy of collapse\footnote{This is
important since Conservation of Mass-Energy demands that the rest mass M of
the object as measured by a distant observer must be the same at the end of
the collapse as it was at the beginning if no energy is radiated away during
the collapse.}.  Once those forces stop the collapse, the object will now
find that it can go to a lower self-gravitational potential energy by
expanding to a larger size, and it will rebound.  The kinetic energy will
then be in place again, but it will be sending the object's mass outward
instead of inward.  The object may then oscillate, but will be losing energy
as it does so.  Once enough energy is lost, it will settle into a stable
state with an average Newtonian potential of $\quantity{\Phi} =
-\quantity{c}^2 / 3$.

\subsection{Hyperstars}
\label{sec:hyperstars}

If blacks holes can not form, then what is a gravitationally collapsed object
like?  First of all, it can not end up having its surface at a radial
coordinate of $r' < 3m$.  If the lowest self-potential energy of a stable
object occurs at $r'=3m$, then the mass of a fully gravitationally collapsed
object to will be distributed around the corresponding potential of
$\quantity{\Phi} = \quantity{c}^2 / 3$ and the surface of such an object
will always be at $r' > 3m$.

This observation leads to the possibility that millisecond pulsars are
neutron stars which are too massive to be stabilized by nuclear/quantum
effects and have undergone runaway gravitational collapse.  However, since
the Schwarzschild radius increases as a function of mass, while volume
increases with the cube of an object's radius, the average density of a
gravitationally collapsed object decreases as its mass increases.  So the
heaviest of the pulsars may not be any more dense than those stabilized by
quantum effects, and may therefore appear to be ``normal'' pulsars.

For heavier objects, the average density decreases still further until at 50
million solar masses you have an object with a Schwarzschild radius of
approximately 1 A.U., and a physical radius of $>$ 1.5 A.U.  Its average
density would be comparable to that of the Sun, and it could be expected to
burn hydrogen like a star.  However its hydrogen burning would
be occurring in a much bigger volume than is the case for a normal star.
Such an object is referred to as hyper-massive star-like object, or
hyperstar.  Hyperstars must generate and emit tremendous amounts of energy,
just as the accretion disks surrounding super-massive black holes are
expected to to do in MGR.  It is therefore speculated that all of the
existing black hole candidates are in fact hyperstars, including the quasars.

\section*{Conclusions}

In this article, it has been demonstrated that it is possible to construct a
general theory of relativity when Extended Uniformity is used instead of
Construction Uniformity.  In the process, it has been shown that if
PGR is true:
\begin{list}{$\bullet$}{\setlength{\topsep}{0ex plus0.2ex}
			\setlength{\parsep}{0ex plus0.2ex}
			\setlength{\itemsep}{0ex plus0.2ex}}
\item
$c$ varies as a function of the observer's GEPUM.
\item
The main predictions of Einstein's MGR are preserved.
\item
The singularities of the Schwarzschild solution are rendered unreachable.
\item
Gravitational collapse produces hyperstars instead of black holes.
\end{list}

At the same time, there are many questions left to answer, such as:
\begin{list}{$\bullet$}{\setlength{\topsep}{0ex plus0.2ex}
			\setlength{\parsep}{0ex plus0.2ex}
			\setlength{\itemsep}{0ex plus0.2ex}}
\item
Can subatomic particles be treated as ``raw matter'' which is stabilized at a
gravitational potential of $-c^2/3$ or as close to it QM will allow?
\item
Can a workable theory of quantum gravity be created using Extended Uniformity?
\item
What are the detailed astrophysics of hyperstars?
\item
What becomes of the Big Bang singularity? How is it reinterpreted in PGR,
or does is even need reinterpretation?
\end{list}

\section*{Acknowledgments}
My thanks to Bill Lawson for his criticisms; to Mark Stuckey for his
comments and encouragement; and to the other participants in the Fifth
Midwest Relativity Conference, whose comments, questions, and presentations
acted as a catalyst for the formulation of the concept of Extended Uniformity.

\appendix
\section{Experimentation and Observation with PGR}
\subsection{PGR's Consistency with existing experimental evidence}
\label{app:consistency}

With regards to existing experiments and observations\cite{Will}, PGR and
MGR produce identical results, due to the preservation of Einstein's field
equations.  For example:
\begin{list}{$\bullet$}{\setlength{\topsep}{0ex plus0.2ex}
			\setlength{\parsep}{0ex plus0.2ex}
			\setlength{\itemsep}{0ex plus0.2ex}}
\item
The deflection of light, as measured photographically and by radio
measurements of occultations of the quasar 3C279:  PGR predicts the same
amounts for the deflection of light, as shown in \S\ref{sec:deflection}
\item
Relativistic time dilation, measured using atomic clocks on aircraft and
rockets: This effect is retained as-is in PGR.
\item
The red-shifting of light, measured for light coming from the Sun and
certain stars, and verified using the Mossbaur effect: Both PGR and MGR
predict the same values for the red-shifting of light, although the theories
differ on how the red-shifted wavelength comes to be observed as detailed in
\S\ref{sec:red-shift}.
\item
The precession of the perihelion of an orbit, as observed for the planet
Mercury and of the binary pulsar PSR\ 1913+16: These effects are obtained
directly from Einstein's field equations, and as such are not affected by
the differences between MGR and PGR.
\item
Radar ranging in the Solar System: Under PGR, the results are interpreted as
representing an expansion of space caused by the presence of the Sun instead
of a decrease in the rate of the propagation of light.  Thus the light is
traveling a longer distance at constant speed, and the ``delay'' expected 
remains the same.
\end{list}

\subsection{Experiments that may test PGR}
\label{app:exper}

In spite of the inability of existing experiments to discriminate between
MGR and PGR, a variation of the Michaelson-Morely experiment can.  Take a
laser beam, split it, send its parts down two paths of \emph{substantially
different total travel length} and then recombine them, producing an
interference pattern.\footnote{This experiment could not be done before the
invention of the laser since incoherent light cannot produce an interference
pattern with total travel paths of substantially different lengths.}  The
experiment then has its GEPUM changed by either gradually moving it to
another elevation above the Earth's surface or keeping it in
place\footnote{The LIGO and VIRGO evacuated tunnels may be ideal places to
do the fixed version of this experiment.} and looking for changes caused by
tidal potential and possibly changes in the Earth's GEPUM caused by
interactions with the Moon and the other planets.

MGR predicts that $c$ as measured at the experiment will be unaffected by
changes in the experiment's GEPUM, which means that the interference pattern
will be unaffected.  PGR predicts that changes in the experiment's GEPUM
will cause $c$ as measured at the experiment to change, resulting in the
relative phases of the light coming in from the two paths to change and the
interference pattern to be modified as a result.  The effects being search
for are quite small, amounting to 1 part in $10^{13}$ for a change of 1
kilometer in the elevation of the experiment.  The experiment will therefore
need to be quite sensitive.  Even so, it should be possible to detect the
PGR effects if PGR is indeed physically correct.

Examining the propagation of light along a standard measuring-rod with
another GEPUM provides another way of testing PGR.  In MGR, all local
observers will measure the same amount of time $t$ for a beam of light to go
back and forth along a standard measuring-rod at their position in a
gravitational field.  As a result, if the movement of light along a standard
measuring-rod with another GEPUM is observed, a back-and-forth time of $t' =
t \lengtheffect$ (due to time dilation) is expected in MGR.  On the other
hand, in PGR the rate of propagation of light with respect to our coordinate
system stays constant while the size of the measuring-rod is decreased from
$l$ to $l \lengtheffect^{-1}$.  This produces the prediction that the
back-and-forth time observed for the measuring-rod with another GEPUM will
be $t' = t \lengtheffect^{-1}$.  Therefore, timing the movement of light
along a standard measuring-rod with another GEPUM will indicate whether
MGR or PGR is physically correct.

Another experiment is the precise and continuous measurement of the time it
takes for light to go back and forth in an evacuated tube, thereby directly
measuring $c$, and detecting variations in it caused by changes in the
experiment's GEPUM.

The fine-structure factor effect given in Eq.\ (\ref{fine-struc-eff})
provides a means of doing an observational test of PGR by examining the fine
structure factor in the emission lines from a gravitationally collapsed
object such as a neutron star or quasar/hyperstar.  For example, suppose
that the surface of a quasar exists at a Newtonian potential of
$\quantity{\Phi} = \quantity{c}^2 / 6$.  In this case, a fine structure
factor of $\alpha' = 2\alpha / 3$ will be observed in the quasar's light,
where $\alpha$ is the fine structure factor for an atom at the potential of
a distant observer.

\subsection{Evidence for the existence of hyperstars}

Evidence for the existence of hyperstars comes from current observations of
galaxies and quasars.  In MGR, quasars are considered to be the accretion
disks of super-massive galaxies which are collapsing into a back hole.  This
view gets some support from the finding that many quasars are accompanied by
galaxies.  However, quasars are known to be much more common in the early
universe \cite{Boyle}, while for more recent times one finds more in the way
of active galaxies, which are believed to have quasar-like cores, prompting
some researchers to consider these as being quasar remnants \cite{Vagnetti}.
This creates the impression that quasars are the progenitors of galaxies
instead of the other way around, which implies that they emit matter instead
of absorbing it as a black hole would do.  In addition, quasars and the
quasar-like central objects of galaxies are known to be relatively small
objects, and the central objects of some galaxies have been shown to have
masses as large as 10 billion solar masses \cite{Ford,Miyoushi}.  These
observations can be interpreted as indicating that quasars and the central
objects of galaxies may represent some condition whereby runaway
gravitational collapse fails to form a black hole.

Another piece of evidence is the recent discovery of carbon in the early
universe \cite{Yamada}.  This may be material formed inside the hyperstar
and later ejected into inter-galactic space.

\begin{table}
\caption{The effects changes in an observer's GEPUM on the measured values of
physical quantities (called observer-based effects), based on the general
relationship $q' = q\lengtheffect^f$.
\label{table:effects}}
\begin{tabular}{llll}
 & &  \multicolumn{2}{c}{Observer-based Effects} \\
\cline{3-4}
Quantity & Units\tablenote{$d$ = Distance, $t$ = Time, $m$ = Mass, $C$ =
Charge} & Mass & Energy \\
& & Fundamental & Fundamental \\ \hline \hline
Length & $d$ & $d' = d\lengtheffect$ & $d' = d\lengtheffect$ \\
Time & $t$ & $t' = t\lengtheffect^{-1}$ & $t' = t\lengtheffect^{-1}$ \\
Velocity & $d/t$ & $v' = v\lengtheffect^2$ & $v' = v\lengtheffect^2$  \\
Acceleration & $d/t^2$ & $a' = a\lengtheffect^3$ & $a' = a\lengtheffect^3$ \\
Mass & $m$ & $m' = m\lengtheffect$ & $m' = m\lengtheffect^{-3}$ \\
Force & $md/t^2$ & $F' = F\lengtheffect^4$ & $F' = F$ \\
Momentum & $md/t$ & $p' = p\lengtheffect^3$ & $p' = p\lengtheffect^{-1}$ \\
Energy & $md^2/t^2$ & $E' = E\lengtheffect^5$ & $E' = E\lengtheffect$ \\
Angular Momentum & $md^2/t$ & $o' = o\lengtheffect^4$ & $o '= o$ \\
\\ \hline
Gravitational Constant & $d^3/mt^2$ & $G' = G\lengtheffect^4$ & 
$G' = G\lengtheffect^8$ \\
Gravitational Potential & $d^2/t^2$ & $\Phi ' = \Phi \lengtheffect^4$ & 
$\Phi ' = \Phi \lengtheffect^4$ \\
GEPUM & $d^2/t^2$ & $\Phi ' = \Phi \lengtheffect^4$ & 
$\Phi ' = \Phi \lengtheffect^4$ \\
\\ \hline
Charge & $C$ & $C' = C\lengtheffect$ & $C' = C\lengtheffect$ \\
Current & $C/t$ & $J = J\lengtheffect^2$ & $J = J\lengtheffect^2$ \\
Electrical Permeability & $C^2t^2/md^3$ & 
$\epsilon_0' = \epsilon_0 \lengtheffect^{-4}$ & $\epsilon_0' = \epsilon_0$ \\
Magnetic Permittivity & $md/C^2$ & $\mu_0' = \mu_0$ & 
$\mu_0' = \mu_0 \lengtheffect^{-4}$
\end{tabular}
\end{table}

\end{document}